\documentclass{vietnam}

\usepackage{amsmath}
\usepackage{xspace}
\usepackage{caption}
\newcommand{\fig}[1]{Fig.~\ref{#1}}

\newcommand{\be}{\begin{equation}}
\newcommand{\ee}{\end{equation}}
\newcommand{\bea}{\begin{eqnarray}}
\newcommand{\eea}{\end{eqnarray}}
\newcommand{\beqn}{\begin{eqnarray}}
\newcommand{\eeqn}{\end{eqnarray}}
\newcommand{\bal}{\begin{align}}
\newcommand{\eal}{\end{align}}
\newcommand{\bitem}{\begin{itemize}}
\newcommand{\eitem}{\end{itemize}}

\newcommand{\qqwwz}{q\bar{q}\to W^+W^- Z}

\newcommand{\PV}{Passarino-Veltman reduction}
\def\gev{\textrm{GeV}}
\def\tev{\textrm{TeV}}
\def\al{\alpha}
\newcommand{\Photo}{}

\begin{document}
\vspace*{0.1cm}
\rightline{KA-TP-34-2013}
\rightline{SFB/CPP-13-79}
\vspace*{4cm}
\title{NLO  $WWZ$ PRODUCTION AT THE LHC}

\author{DAO Thi Nhung$^{1,2,}$\footnote{Speaker}, LE Duc Ninh$^{1,2}$ and Marcus M. WEBER$^3$}

\address{{\it $^1$Institut f\"ur Theoretische Physik, Karlsruher Institut f\"ur  Technologie, \\
D-76128 Karlsruhe, Germany\\
$^2$ Institute of Physics, Vietnam Academy of Science and Technology, \\
10 Dao Tan, Ba Dinh, Hanoi, Vietnam\\
 $^3$Max-Planck-Institut f\"ur Physik (Werner-Heisenberg-Institut), \\
D-80805 M\"unchen, Germany}}

\maketitle\abstracts{The tri-boson production is one of the key processes for the study of quartic gauge couplings. Next-to-leading
order (NLO) corrections are mandatory to reduce theoretical uncertainties. In this study, 
the most up-to-date predictions including NLO QCD and NLO EW corrections to the total cross section and distributions of the
 $W^+W^-Z$ production at the LHC are presented. We show that the QCD correction is about  $100\%$ and
the EW correction is of  a few percent  at the total cross section level. 
The EW correction however
becomes significant in the high energy regime of the gauge boson transverse momentum distributions.}

\section{Introduction}

In the Standard Model (SM), electromagnetic and weak interactions  are  described  
by the local gauge group 
$SU(2)_L\otimes U(1)_Y$. According to  this non-Abelian gauge structure, there exists  three-
and four-gauge-boson interactions. While the triple gauge couplings ($WW\gamma, WWZ$) 
 are experimentally well understood through the
analysis of two-gauge-boson production at the LEP, Tevatron and LHC (7 and 8 TeV), 
the understanding of  quartic 
gauge couplings ($WW\gamma\gamma, WW\gamma Z, WWZZ, WWWW$) is less advanced.
This limitation is due to the limited energy of those experiments. The direct signals of these quartic gauge couplings 
 are obtained by either di-boson production in association with two fermions or tri-boson production
which of course require high center-of-mass energy. In the next few years, operation of the upgraded LHC at 13 and 14 TeV
may provide enough sensitivity for the study of the two mechanisms.

In this talk, we present  a detailed study of the  $pp\to W^+W^-Z$ process which is sensitive 
to the  $WW\gamma Z$ and $ WWZZ$ couplings. This process is also a background to the
 search of beyond SM physics. We focus on the quantum corrections to the total cross section
and distributions of the process, particularly the NLO QCD and electroweak (EW) corrections.
While the NLO QCD correction has been studied in Ref~\cite{Hankele:2007sb,Binoth:2008kt}, the NLO EW correction  has been recently computed 
by our group~\cite{Nhung:2013jta}. We recomputed also the NLO QCD correction and combined with the NLO EW one to provide
the most  up-to-date prediction for this process.
%%%%%%%%%%%%%%%%%%%%%%%%%%%%%%%%%%%%%%
\section{Calculation}
In this section we summarize the main points of our calculation, for  more details we refer  to Ref~\cite{Nhung:2013jta}.
\begin{figure}[h]
  \centering
 \includegraphics[width=0.8\textwidth,height=0.15\textwidth]
{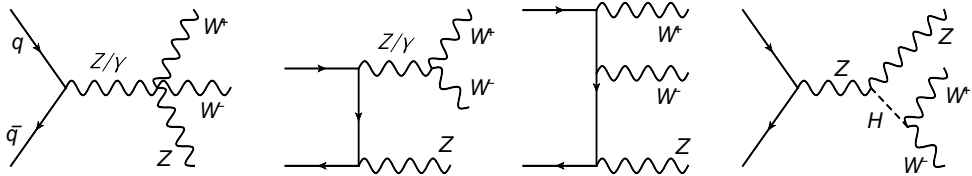}\\
  \caption{ Representative tree-level diagrams for the
$\qqwwz$ subprocesses.}\label{qqWWZ_tree_diag}
\end{figure}
The tree-level subprocesses are
\be
a)~~\bar{q} + q  \to  W^+ + W^- + Z,\quad b)~~ 
\bar{b} + b  \to  W^+ + W^- + Z, \quad c)~~
\gamma + \gamma  \to  W^+ + W^- + Z,
\ee
where $q$ stands for the light quarks ($u,d,c,s$) if not otherwise stated.
The $q\bar{q}$ contributions, whose Feynman diagrams are depicted in \fig{qqWWZ_tree_diag}, are dominant. We include the $s$-channel diagrams with an intermediate Higgs boson.
The Higgs contribution including interference effects is less than $1\%$ at leading order (LO) for
$M_H = 125\;\gev$.  The LO $b\bar{b}$ contribution is  2.4\% while the LO $\gamma\gamma$ contribution is  about 5\% 
compared to the LO light quark cross section.
We therefore  compute NLO corrections to the light quark contribution, but not for the $b\bar{b}$ and $\gamma\gamma$ contributions.

The NLO QCD correction to $\qqwwz$ consists of virtual and real  corrections at the ${\cal O}(\al_s\al^3)$ order. The UV
divergences appearing in the virtual part are isolated using dimensional regularization. The soft and collinear
singularities occur in the initial state radiation of gluon (quarks) and in the virtual contribution. 
We have employed two methods,
dimensional  and mass regularizations, to separate IR singularities. The results obtained by both methods
are in good agreement. To combine the virtual and real  contributions we apply the  dipole subtraction algorithm
 by Catani-Seymour~\cite{Catani:1996vz}. The soft singularities are canceled out in the combined contribution and
the left-over collinear singularities are absorbed in the quark distribution functions.  
To have a better understanding of the origin of large QCD corrections, we
divide the QCD correction into virtual, gluon radiated and gluon-quark induced contributions. The virtual contribution
consists of the QCD-loop contribution and the I-operator part arising from real radiation as defined in Ref~\cite{Nhung:2013jta}. The gluon radiated contribution
is the remaining part of the real gluon emission process after subtracting the I-operator part. 

The computation of the NLO EW correction is much more complicated in comparison with the NLO QCD correction because of
the involvement of many EW particles $W, Z, \gamma, H$ and fermions. The NLO EW correction contains both one-loop EW and real photonic
contributions. To get the UV-finite results of the virtual part, the on-shell (OS) renormalization scheme for  boson masses $M_W, M_Z, M_H$
and external particle wave functions is employed. The $G_\mu$ scheme is applied for the electric charged $e$. The NLO EW contribution contains 
a real photon whose coupling must be defined in the Thomson limit. We therefore have to rescale the NLO EW correction with a factor of $\al_{0}/\al_{G_\mu}$. 
  Similar to the NLO QCD correction,
we also divide the NLO EW corrections into the virtual, photon radiated and photon-quark induced contributions. Each contribution alone is free
of UV and IR divergences. Mass regularization is used in combination with the dipole subtraction method~\cite{Dittmaier:1999mb}  to deal with
IR divergences. 

The  matrix elements are generated with the help of {\texttt{FeynArts-3.4}}~\cite{Hahn:2000kx} and
{\texttt{FormCalc-6.0}}~\cite{Hahn:1998yk} as well as
{\texttt{HELAS}}~\cite{Murayama:1992gi,Alwall:2007st}.
The maximum tensor loop integrals encountered in both  QCD and EW virtual parts are the five-point tensor integrals of rank 3. The traditional \PV~\cite{Passarino:1978jh}
are used for tensor reduction. 
The scalar and tensor
one-loop integrals in one code are evaluated with the in-house library {\texttt{LoopInts}}.  
The library automatically uses  quadruple precision when it detects  a small  Gram determinant of the $N$-point
 tensor coefficients ($N=3,4$). Otherwise it uses double precision.   
\section{Results}
In this section we highlight the important numerical results for LHC 14 TeV using the MSTW2008 parton distribution function (PDF) set,
 for a more detailed discussion we refer  to Ref~\cite{Nhung:2013jta}.

\begin{figure}[h]
\begin{minipage}{0.6\textwidth}
~~~~We start with a discussion of scale dependence. The LO total cross section including only the $q\bar q$ contribution depends solely on 
the factorization scale $\mu_F$ entering through  PDFs. The NLO QCD results introduce dependence on renormalization scale $\mu_R$ through $\al_s$. 
The NLO EW correction does not depend on $\mu_R$ since the OS renormalization scheme is used. 
 For simplicity, we set $\mu_F=\mu_R=\mu$.      
In \fig{scale} we show the total cross sections and K-factor as functions of $\mu$ varied
around the center scale $\mu_0$ for two cases: a fixed scale with
$\mu_0=2M_W+M_Z$ and a dynamic
scale $\mu_0=M_{WWZ}$, the invariant mass of the tri-boson system. The K-factor is defined as the ratio of the
NLO QCD cross section with respect to the LO one. The strong dependence of the NLO QCD cross section on the
scale is traced to the dependence of $\al_s$ on $\mu_R$. The dynamic scale results are similar to the
fixed ones for both the total cross section and the distributions we have studied. We chose the fixed scale
at $\mu=2M_W+M_Z$ for the rest of discussion.\end{minipage} 
\begin{minipage}[l]{0.40\textwidth}
%\begin{figure}
\captionsetup{width=0.8\textwidth}
 \begin{center}
\includegraphics[width=1\textwidth]{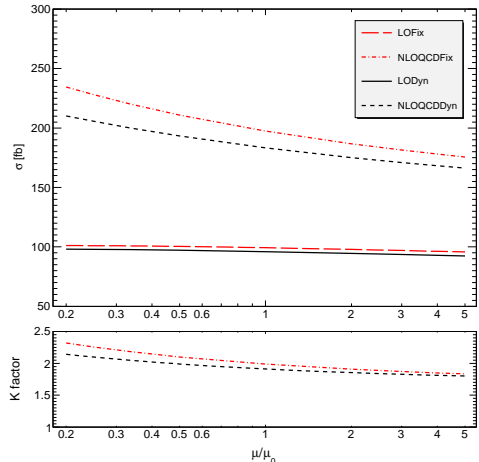}
\caption[width=0.4\textwidth]{The total cross sections and K-factors as functions of scale.}
\label{scale}
\end{center}
%\end{figure}
\end{minipage}
\end{figure}

\begin{figure}[h]
 \begin{center}
\includegraphics[width=0.32\textwidth]{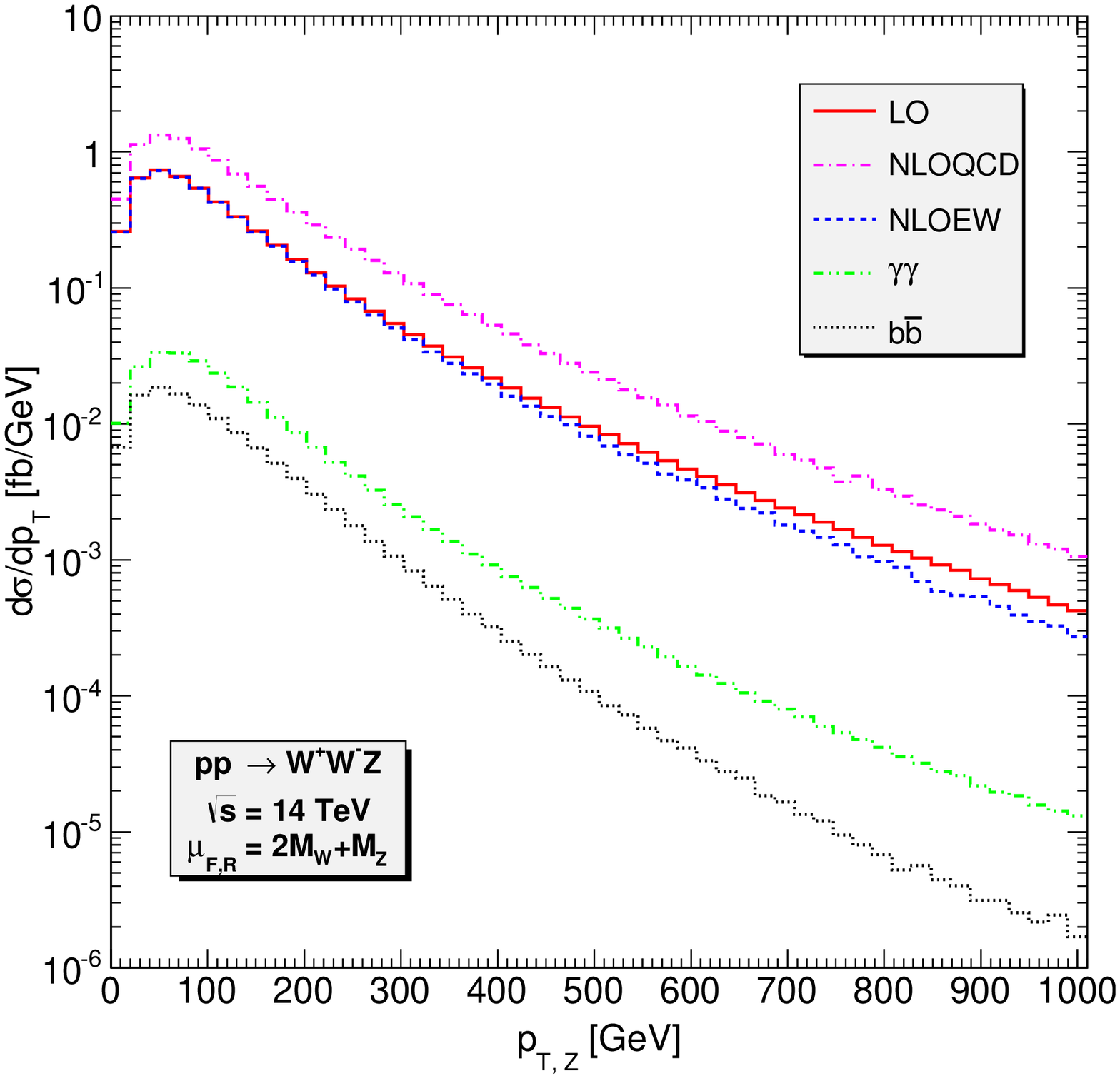}
\includegraphics[width=0.32\textwidth]{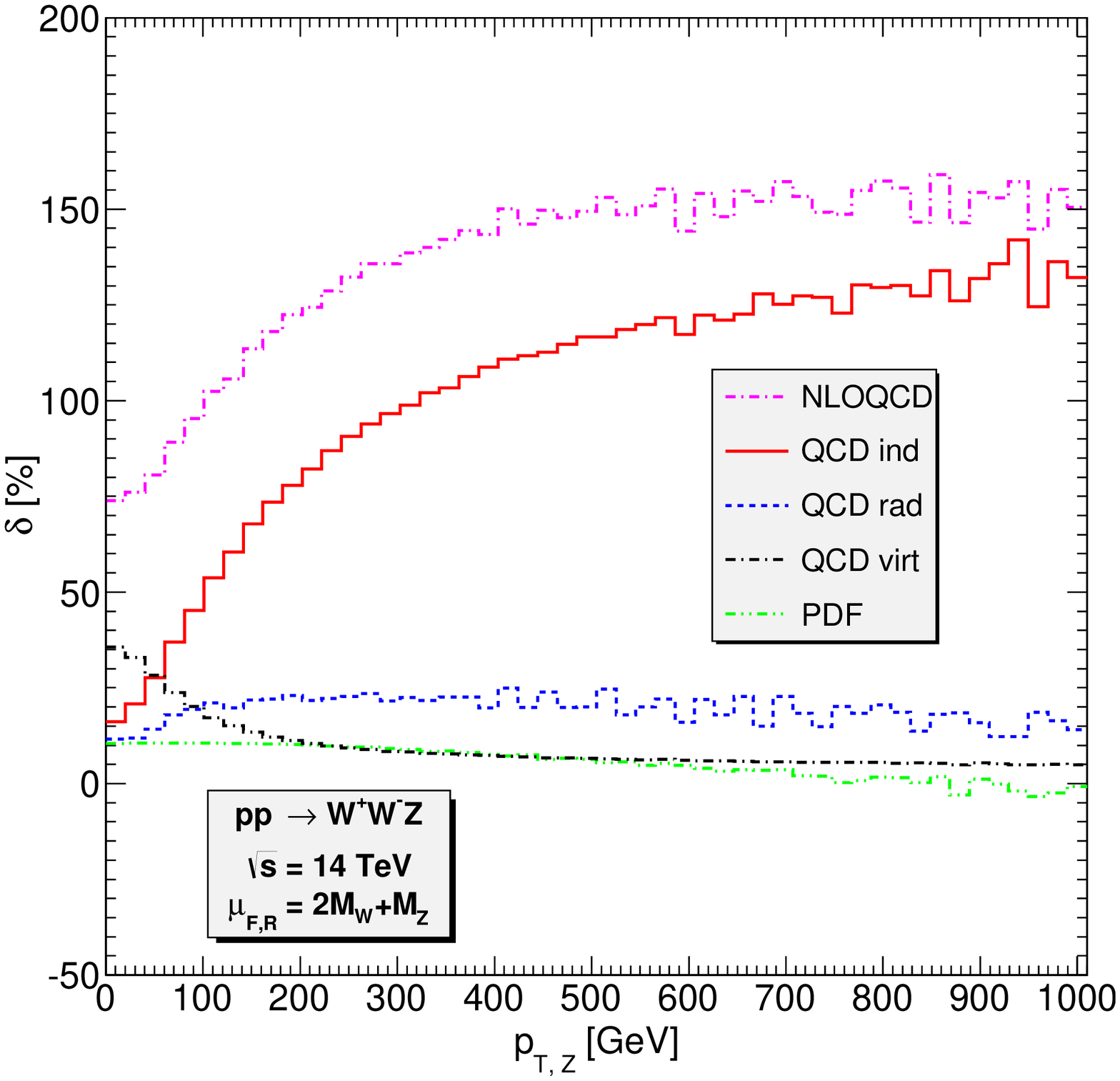}
\includegraphics[width=0.32\textwidth]{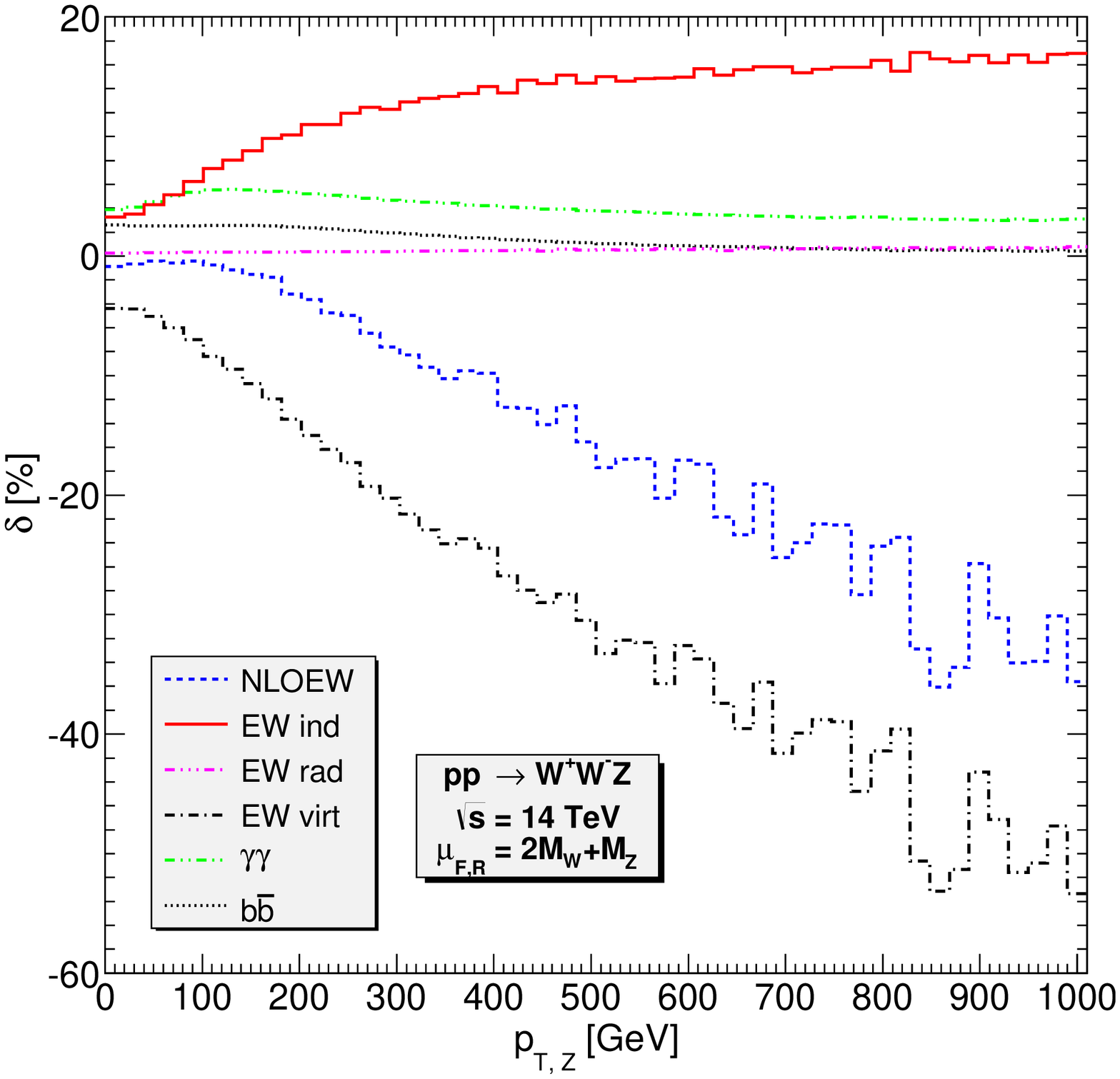}
\caption{$Z$ transverse momentum distribution of $pp \to W^+W^-Z$ cross section (left),
of the NLO QCD corrections (middle) and of the NLO EW corrections (right).
}
\label{qqWWZ_pt}
\end{center}
\end{figure}

We present the $p_T$ distributions for the LO contribution as well as the
$b\bar b$, $\gamma\gamma$, NLO QCD and NLO EW corrections in the left panel of \fig{qqWWZ_pt}. We take the $p_T$ 
of the Z-boson as an example. The separated NLO QCD and NLO EW
relative corrections compared to the LO distributions are also shown in the middle and right panels of \fig{qqWWZ_pt}, respectively.
The $b\bar b$ and $\gamma\gamma$ contributions are about 1 to 2 orders of magnitude smaller
than the $q\bar q$ contribution in the whole $p_T$ range. The NLO QCD correction increases rapidly in the low $p_T$ range
and is nearly constant for $p_{T}>400\;\gev$. The dominant contribution comes from the
gluon-induced subprocesses.
The remaining contributions are less than $30\%$. For the NLO EW corrections, the virtual part is negative in the whole $p_T$ range and behaves like $\alpha\log^2(M_V^2/p_T^2)$,
reaching about $-50\%$ at $p_T = 1\;\tev$.
This is the well-known Sudakov double logarithm arising from the exchange of a virtual massive
gauge boson in the loops. The photon-induced
correction is about $+20\%$ at $p_{T,Z}=1\;\tev$, canceling part of the Sudakov virtual correction.

\begin{figure}[h]
 \begin{center}
\includegraphics[width=0.32\textwidth]{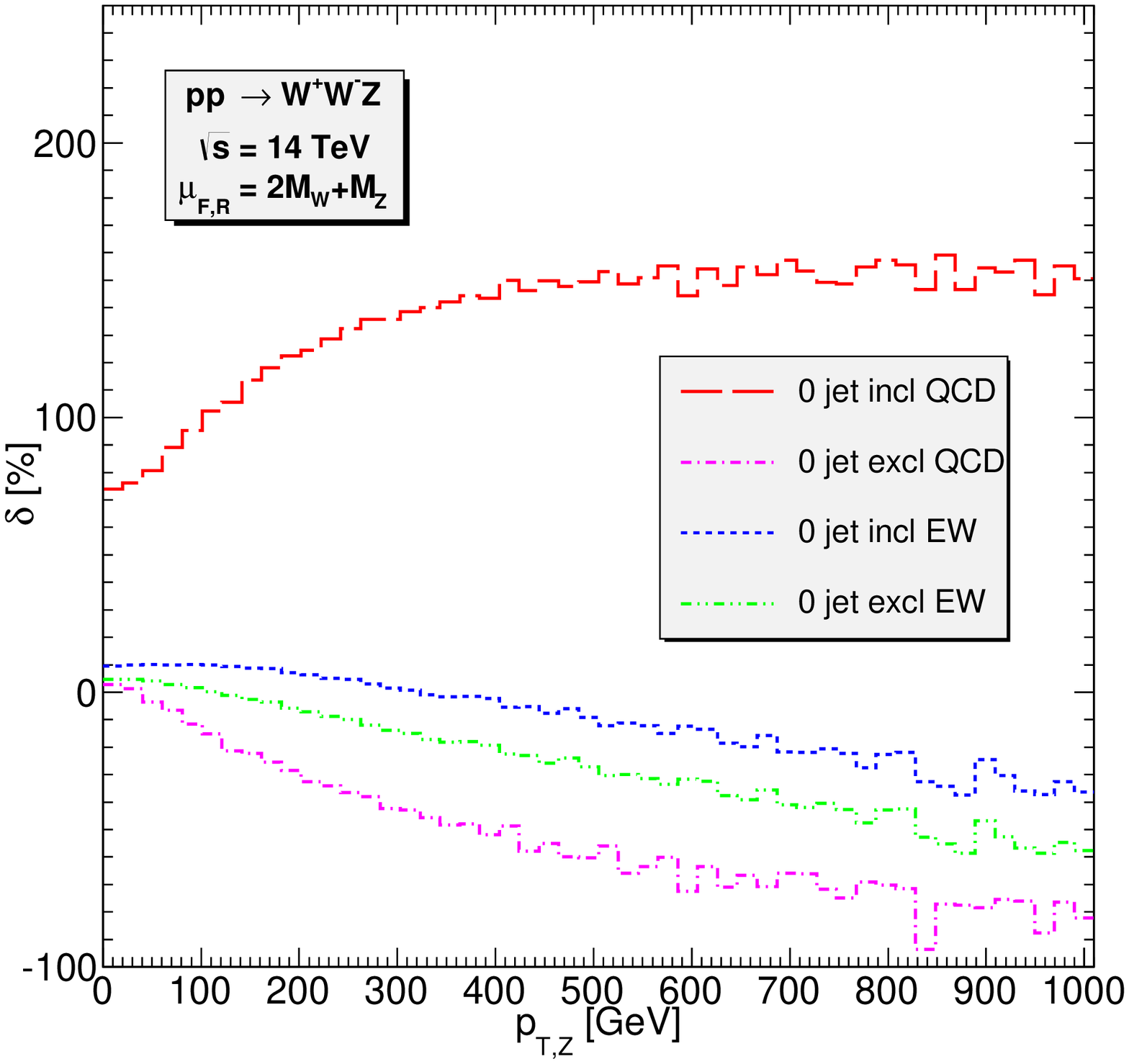}
\includegraphics[width=0.32\textwidth]{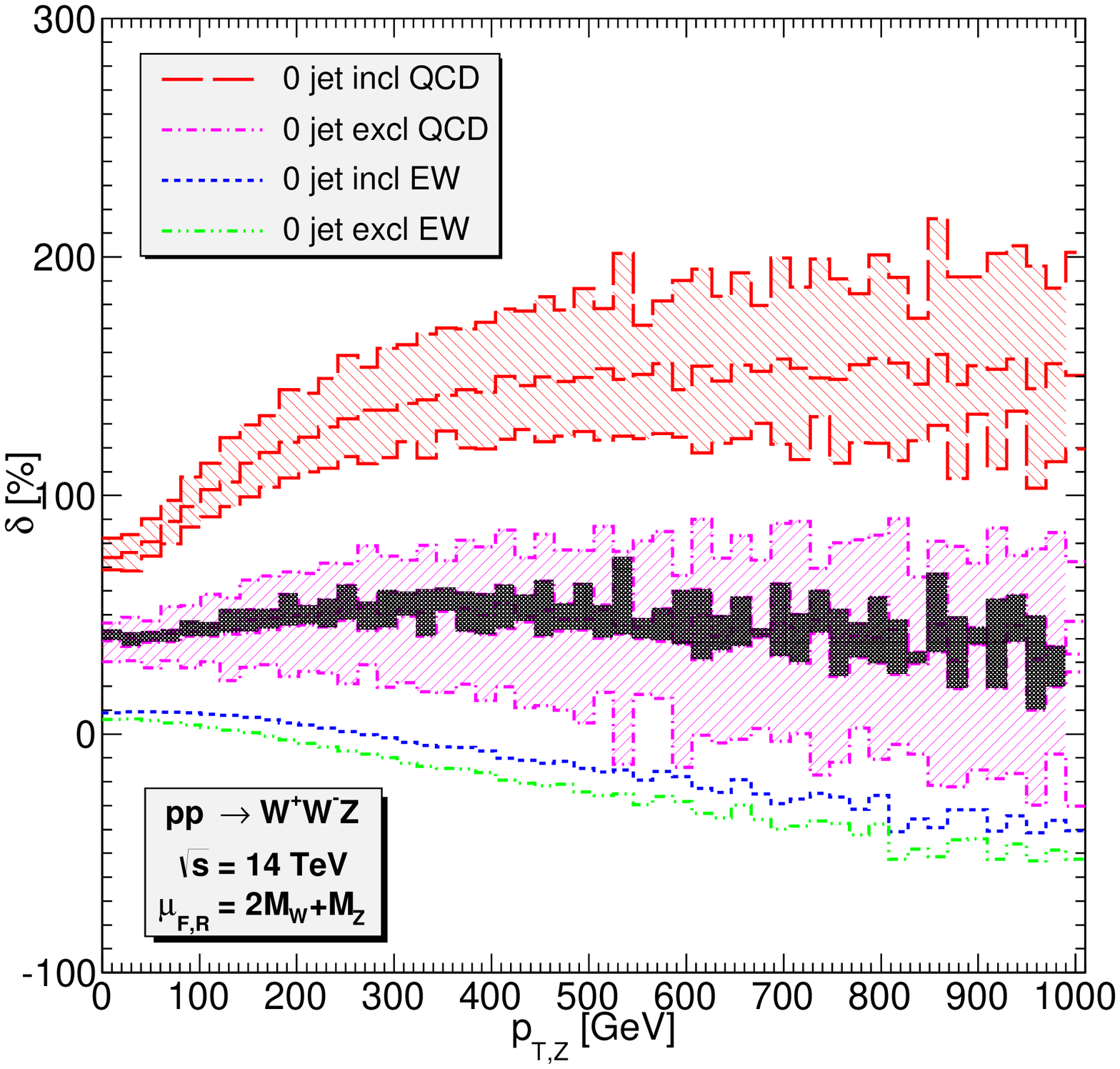}
\caption{NLO QCD and EW corrections to the $Z$ transverse momentum distribution for
inclusive events without jet cuts and also for exclusive events with  fixed jet veto (left) and  dynamic
jet veto with uncertainty bands (right).
}\label{fig:jetveto_dyn}
\end{center}
\end{figure}

From the above phase-space dependence study, we see that the NLO QCD correction mainly due to the
$2 \to 4$ gluon-quark induced channels is positive and very large at high $p_T$. The NLO EW is negative and mainly comes 
from the $2 \to 3$  virtual correction. 
One therefore thinks of
imposing a jet veto to reduce this large QCD contribution.
We have tried a fixed jet veto with $p_\text{veto} = 25\;\gev$ and
found that it over subtracts the NLO QCD correction,
leading to a large negative QCD correction at high $p_{T,Z}$, see left plot of \fig{fig:jetveto_dyn}.
 The situation is better with a dynamic jet veto with
$p_\text{veto} = 1/2 \text{max}(M_{\text{T},W^+},M_{\text{T},W^-},M_{\text{T},Z}),$
where $M_{\text{T},V}=(p_{\text{T},V}^2 + M_{V}^2)^{1/2}$ is the transverse mass. We found that
more than half of the QCD correction is removed. However, using a jet veto increases theoretical
uncertainty due to missing large higher-order corrections which is
supported in the right plot of \fig{fig:jetveto_dyn}. The uncertainty band on the
exclusive zero-jet distribution (in pink) is larger than the band on the inclusive zero-jet distribution. 
Note that, the  uncertainty band (in black) is severely underestimated when the zero-jet and one-jet inclusive observables
are wrongly assumed be anti-correlated.  
The bands describe $\mu_0/2 \le \mu_F = \mu_R \le 2\mu_0$ with $\mu_0 = 2M_W+M_Z$ variations 
of the NLO QCD corrections.
%%%%%%%%%%%%%%%%%%%%%%%%%%%%%%%%%%
\vspace*{-10pt}
\section{Conclusions}
In this talk, we have discussed  the  NLO EW and  NLO QCD corrections
to the $W^+W^-Z$ production at the LHC at $14\;\tev$ center-of-mass energy.
We discussed also the use of  a jet veto to reduce the large QCD correction. 
\section*{Acknowledgments}

D.T.N thanks the organizers for the nice atmosphere during the conference and the possibility
to give  talk. This work is supported by the Deutsche
Forschungsgemeinschaft via the 
Sonderforschungsbereich/Transregio 
SFB/TR-9 Computational Particle Physics.  
\section*{References}

\end{document}